**Nonmagnetic ground state of marcasite $FeTe_2$: The competition between crystal field splitting and on-site Coulomb repulsion**

Yue-Fei Hou,[1,2,6] Zhibin Shao,[3] Minghu Pan,[4,*] Shiyang Wu,[5] Fawei Zheng,[5] Zhen-Guo Fu,[1,6,‡] and Ping Zhang,[1,6,7,†]

[1]Institute of Applied Physics and Computational Mathematics, Beijing 100088, China
[2]Graduate School, China Academy of Engineering Physics, Beijing 100088, China
[3]Physics Laboratory, Industrial Training Center, Shenzhen Polytechnic University, Shenzhen 518055, China
[4]School of Physics and Information Technology, Shaanxi Normal University, Xi'an 710119, China
[5]School of Physics, Beijing Institute of Technology, Beijing 100081, China
[6]National Key Laboratory of Computational Physics, Beijing 100088, China
[7]School of Physics and Physical Engineering, Qufu Normal University, Qufu 273165, China

## ABSTRACT

The magnetic ground states in crystalline systems are significant for both fundamental condensed matter physics and practical materials engineering. Marcasite $FeTe_2$, characterized as a small-gap semiconductor, exhibits anomalous magnetic behaviors in low-temperature experiments. In this study, first-principles density functional theory calculations combined with scanning tunneling microscopy/spectroscopy are employed to investigate the magnetic ground state of marcasite $FeTe_2$. It is revealed that the competition between crystal field splitting and on-site Coulomb repulsion plays the key role in the formation of localized magnetic moments in $FeTe_2$. The ground state of $FeTe_2$ bulk is confirmed to be nonmagnetic, while the magnetic responses of $FeTe_2$ observed at low temperature are suggested to be related to the magnetic Fe atoms on the crystal surfaces. Our work proposes a straightforward competing mechanism for determining ground-state magnetism of various localized-moment crystalline systems.

[*]Contact author: minghupan@snnu.edu.cn

[‡]Contact author: fu_zhenguo@iapcm.ac.cn

[†]Contact author: zhang_ping@iapcm.ac.cn

## I. INTRODUCTION

Transition metal dichalcogenides have attracted considerable research interest owing to their rich physical properties and promising applications in diverse fields such as spintronics, catalysis, energy storage, and optoelectronics [1-5]. Within this family of materials, $FeTe_2$ exhibits unique structural versatility, crystallizing in either van-der-Waals layered structures or fully bonded bulk structures (marcasite- or pyrite-type) [6,7]. Recent advances have enabled the synthesis of three-dimensional layered $FeTe_2$ nanocrystals with the hexagonal lattice, which display semiconducting characteristics [8]. Intriguingly, theoretical predictions for the single-layer $FeTe_2$ suggest an intrinsic metallic electronic structure [9] and potential room-temperature ferromagnetic ordering [10,11], though these properties await experimental verification. In contrast to the hexagonal layered structure, orthorhombic marcasite $FeTe_2$ has been extensively investigated since last century, primarily for its thermoelectric performance [7,12-15]. Both theoretical and experimental studies consistently identify marcasite $FeTe_2$ as a small-gap semiconductor [13-22]; however, the exact nature (direct versus indirect) and the width of its bandgap remain subjects of ongoing investigation. Notably, marcasite $FeTe_2$ exhibits a wealth of intriguing magnetic phenomena, including room-temperature magnetic ordering [19-21,24,25], antiferromagnetic coupling between magnetic moments [21-23,25], multiple magnetic phase transitions across temperature regimes [21,25], and significant magnetic anisotropy [20,21]. This unique combination of semiconductivity and complex magnetic properties positions marcasite $FeTe_2$ as a particularly promising candidate for spintronic applications, meriting further in-depth exploration.

Nevertheless, the magnetic properties of marcasite $FeTe_2$ in its ground state remain controversial across different experimental studies that extend measurements to near-zero temperatures. Magnetic susceptibility measurements [22], along with zero-field-cooling (ZFC) and field-cooling (FC) curves [23, 25], provide evidence supporting an antiferromagnetic (AFM) ground state. However, another magnetization combined with resistivity measurements [21] suggest instead a ferromagnetic (FM) ground state. Furthermore, extrapolations of magnetization to zero temperature yield inconsistent results among different research groups [20, 21, 23, 25], all indicating an unsaturated localized magnetic moment. Given that these studies indeed synthesized the consistent crystal structure, the origin of these discrepancies in magnetic measurements remains unclear. Additionally, systematic theoretical calculations to determine the ground-state magnetism of marcasite $FeTe_2$ are still lacking. Resolving this issue is essential not only for fundamental physics but also for potential applications. Thus, a clear identification of the ground-state magnetic properties of $FeTe_2$, supported by robust theoretical explanations, is critically needed.

In this work, we performed first-principles density functional theory (DFT) calculations as well as scanning tunneling microscopy/spectroscopy (STM/STS) measurements to reveal the ground state of marcasite $FeTe_2$. In $FeTe_2$ single crystal, the 3*d* electrons are localized and exhibit more orbital-like behaviors, leading to nodal lines or nodal planes of the electronic density. To reasonably describe this non-uniformity, the Perdew-Burke-Ernzerhof (PBE) type generalized gradient approximation (GGA) functional [26] is applied in our DFT calculations. The DFT plus Hubbard's U (DFT+U) method [27] is applied to include the on-site Coulomb interactions. In the experiment, the molecular beam epitaxy (MBE) method is adopted to prepare marcasite $FeTe_2$ film on $SrTiO_3$ substrate in an ultra-high vacuum, followed by annealing under a Te-rich atmosphere. STM/STS is employed to display the atom-resolved bias-dependent images and the electronic density of states around the Fermi level of the MBE-grown $FeTe_2$ film.

Based on these explorations, it is found that whether the Fe atoms in marcasite $FeTe_2$ form localized magnetic moments or not can be directly related to the competition between crystal field (CF) splitting and on-site Coulomb repulsion of the localized 3*d* electrons. The ground state of $FeTe_2$ bulk is confirmed to be nonmagnetic (NM) with a small indirect energy gap, while the surfaces of $FeTe_2$ crystal are spin-polarized with potential two-dimensional (2D) AFM orders. This picture of ground-state $FeTe_2$ not only supports its semiconducting nature, but also explains the variability in measured saturated magnetization [20,21,23,25] (reported in the unit of emu/g or emu/mol) may be caused by different surface-to-volume ratios of different samples. Our work identifies a new example of Fe-based single crystal with completely quenched localized magnetic moments in the bulk, and demonstrates a dominating competing mechanism in determining its localized moments.

## II. RESULTS

### A. The STS gap and the formation of localized magnetic moments

Marcasite $FeTe_2$ has a body-centered orthogonal structure. The space group is *Pnnm* (58). Sample preparation was conducted in a molecular beam epitaxy chamber with a base pressure of $3\times10^{-10}$ Torr. Firstly, the nearly monolayer FeTe films on the $SrTiO_3$ substrate were grown. Subsequently, FeTe films were annealed under a rich-Te atmosphere for 15 minutes to obtain a rich-Te phase of $FeTe_2$ compounds. As revealed in Fig. 1(a), the STM observations show a clean and high-quality 010 surface of $FeTe_2$ without reconstruction of the surface atoms. By measuring the atomic distances on the 010 surface and the step height along the 010 direction of the crystal, the lattice parameters are determined as $a$=3.75 Å, $b$=5.4 Å, and $c$=6.1 Å. These values are in reasonable agreement with the x-ray diffraction (XRD) results [21]. More details of the measured lattice parameters are given in Section A of the Supplemental Material (SM) [28]. As shown in Fig. 1(b), the STS spectrum shows an energy gap for about 140 meV in $FeTe_2$ at the temperature of 4.2 K. At 77 K, the gap persists but slightly narrows to 95 meV due to the thermodynamic effects. The STS results confirm the small-gap semiconductivity of marcasite $FeTe_2$ single crystal.

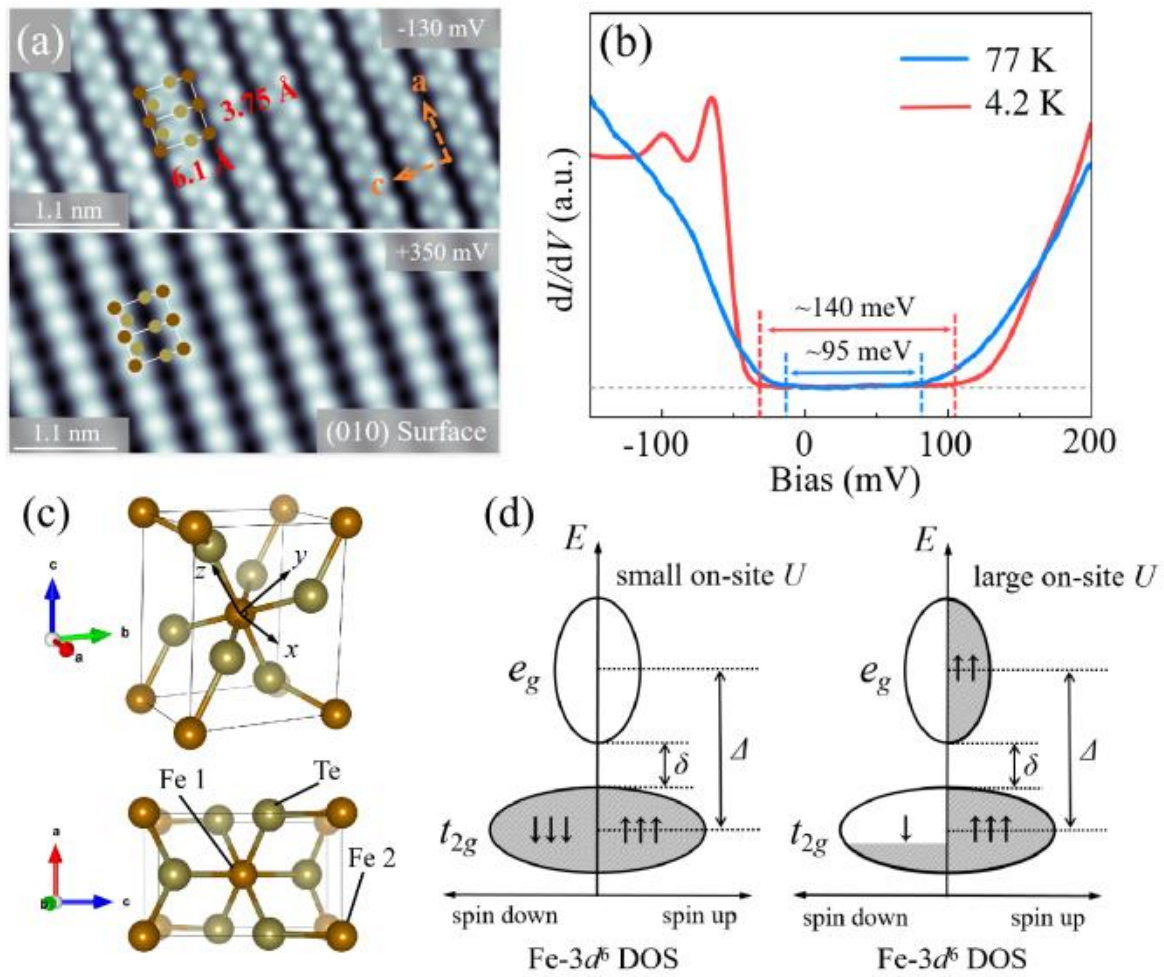

FIG. 1. (a) Atomic resolution images acquired at different biases, visualizing the Te and Fe atomic layers of the 010 surface. Image size and scanning current: $5.6\times2.6$ nm$^2$ and 200 pA. (b) d$I$/d$V$ spectra acquired at 4.2 K and 77 K. The dashed lines schematically mark the range of energy gap. Set point: $V_b$ = 300 mV, $I_t$ = 200 pA. (c) The conventional cell of marcasite $FeTe_2$ single crystal. The Cartesian coordinate system used in the following simulations is shown. (d) The status of spin occupation on Fe-3$d$ states in the cases with a small on-site $U$ and with a large on-site $U$. $\Delta$ denotes the CF splitting, and $\delta$ denotes the energy gap between CF splitting subbands. The renormalization of the electronic density of states (DOS) caused by on-site $U$ from two anti-parallel spins is not explicitly shown for simplicity.

The conventional cell of marcasite $FeTe_2$ is used for DFT simulations. There are two Fe atoms and four Te atoms in the unit cell, as shown in Fig. 1(c). There are six nearest Te ligands around each Fe, forming a distorted $FeTe_6$ octahedral structure. Before performing DFT calculations, a theoretical analysis is given based on the STS gap to provide an expectation for the ground state of $FeTe_2$. For an isolated Fe atom with the $3d^6$ electronic configuration, the total spin is $S$ = 2, as described by Hund's rules. This spin configuration is driven by both exchange and Coulomb interactions. We propose that in the $FeTe_2$ crystal the spin configuration of Fe-3$d$ electrons should be additionally affected by the CF splitting, competing with the on-site Coulomb repulsion to dominate the formation of local magnetic moments. Fig. 1(d) shows the expected spin occupations of Fe-3$d$ electrons in crystal bands, suggesting a NM or a magnetic ordered ground state of $FeTe_2$ with a small on-site $U$ or a large on-site $U$, respectively. In this framework, the one-electron approximation and a large $e_g$-$t_{2g}$ CF splitting ($\Delta$) assumed for the $FeTe_2$ crystal are adopted. With finite widths of the $e_g$ and $t_{2g}$ subbands, a non-zero energy gap $\delta$ should be reserved as recommended by the STS spectrum.

In our practical calculations, Dudarev's DFT+U method [27] with the single adjustable parameter is applied to reasonably describe the on-site Coulomb repulsion:

$$E_{\mathrm{DFT+U}} = E_{\mathrm{DFT}} + \frac{U-J}{2}\sum_{\sigma}\left[\mathrm{Tr}\rho^{\sigma} - \mathrm{Tr}\left(\rho^{\sigma}\rho^{\sigma}\right)\right]$$
$$= \frac{U_{\mathrm{eff}}}{2}\sum_{\sigma}\left[\mathrm{Tr}\rho^{\sigma} - \mathrm{Tr}\left(\rho^{\sigma}\rho^{\sigma}\right)\right].$$

$\rho^{\sigma}$ represents the density matrix of the localized $d$ electrons with spin $\sigma$ ($\sigma$=↑ or ↓). The on-site Coulomb repulsive energy $U$ and orbital exchange energy $J$ are combined in the effective Coulomb repulsive energy $U_{\mathrm{eff}}$. By increasing the value of $U_{\mathrm{eff}}$ in the calculations, the computed properties will vary. This allows us to determine the strength of Coulomb repulsion of 3*d* electrons by comparing the simulation results to the experimental results. More details about the computational methods and parameters are given in Section B of the SM [28]. A detailed discussion about the other non-dominant interactions in $FeTe_2$ that affect the localized magnetic moments is also given in Section C of the SM [28].

### B. The nonmagnetic ground state of $FeTe_2$ bulk

With variable $U_{eff}$ being applied, the simulated physical properties are shown in Fig. 2. As shown in Fig. 2(a), the cell volume increases with increasing $U_{eff}$ for the NM, FM, and AFM states. Although the applied GGA exchange-correlation functional has the potential to predict slightly larger lattice parameters than the real samples [29-31], the cell volumes for the magnetic ordered states (FM 1, FM 2, and AFM) deviate severely from the experimental results. Only the NM state is reasonably consistent with the experiments. For the magnetic ordered states, the spin-paired 3*d* electrons are driven to be polarized and occupy higher-energy CF orbitals, which significantly strengthens the repulsion between the Fe and its Te ligands. The cell volume will thus increase for the magnetic ordered states to release this stress. The detailed data of lattice parameters and local magnetic moments for all the computed states are given in Section D of the SM [28].

By comparing the total energies of different magnetic ordered states, a first-order phase transition from NM to AFM is verified when $U_{eff}$ is about 1.5 eV, as shown in Fig. 2(b). The simulated Fe-3*d* spin configurations of the NM state and the AFM state are reasonably consistent with the prediction in Fig. 1(d) for all the values of $U_{eff}$. However, the increase in the energy difference between the NM state and the AFM state deviates from the linear relationship recommended by Hubbard's *U* model. This phenomenon reflects the variation of the occupation numbers of the 3*d* orbitals with different values of $U_{eff}$, as discussed in Section E of the SM [28].

FIG. 2. Simulated physical properties with different values of $U_{eff}$. (a) The volumes of the conventional cell for the NM, FM, and AFM states. FM 1 and FM 2 are two different FM metastable states [28,34,35] with different 3*d* localized orbital states. The AFM state is simulated by setting the magnetic moments of Fe 1 and Fe 2 in the unit cell to be anti-parallel. The XRD result [21] and our STM results are marked by the black dot line and the red dashed line, respectively. (b) The relative total energies for the NM, FM, and AFM states. The NM-AFM transition occurs at $U_{eff}$ = 1.5 eV. The energy difference between the NM state and the AFM state is denoted by the black solid squares at the bottom right corner. (c) The indirect energy gaps for the NM state and the magnetic ordered states. Our STS gap is marked by the red dashed line.

[*]Contact author: minghupan@snnu.edu.cn

[‡]Contact author: fu_zhenguo@iapcm.ac.cn

[†]Contact author: zhang_ping@iapcm.ac.cn

The existence of an energy gap for a semiconductor should be consistent across both experimental observations and DFT calculations. Our STS measurements indicate an energy gap of 140 meV for $FeTe_2$. Therefore, the widths of the gap for different states are simulated using various $U_{eff}$ values to determine the most reasonable $U_{eff}$. As shown in Fig. 2(c), the NM state exhibits an increasing indirect gap as $U_{eff}$ increases. However, the simulated magnetic ordered states (FM 1, FM 2, and AFM) of $FeTe_2$ all exhibit metallic behavior with no energy gaps. To achieve reasonable consistency with the experimental facts, the value of $U_{eff}$ for the following ground-state simulations is ultimately determined as 1.2 eV, which best describes both the NM and gapped properties of ground-state $FeTe_2$. We noticed that first principles calculations suggest that the value of Hubbard's $U$ be in the range of 2-5 eV [32] ($U_{eff}$ is in the similar range since $J$ is usually an order smaller than $U$), while the measurements of x-ray excited Auger spectra [33] only suggested a $U_{eff}$ of about 1.25 eV for Fe atoms. In essence, our DFT+STM/STS investigations agree with the relatively small value of $U$ ($U_{eff}$) for marcasite $FeTe_2$, although a large $U$ is indeed achieved by a linear response method as shown in Section F of the SM [28]. The electronic structures and phonon structures for the magnetic ordered states are also provided in Section G of the SM [28].

Although both the XRD measurement [21] and our DFT calculations have observed distortions of the $FeTe_6$ octahedral structure in marcasite $FeTe_2$, an $e_g$ - $t_{2g}$ splitting mode should still be partly maintained. In light of the specific distortions in $FeTe_2$, a local Cartesian coordinate system is established to better describe the $e_g$ - $t_{2g}$ splitting mode, as depicted in Fig. 3(a). The simulated lengths of the Fe-Te bonds in the $x$-$y$ plane and the $y$-$z$ plane are 2.57 Å and 2.55 Å, respectively. Due to the rotation of the Fe-Te bonds in the $x$-$y$ plane, the largest Te-Fe-Te angle becomes 98.5°. These simulation results are in good agreement with the XRD results [21], that the two Fe-Te bond lengths are 2.56 Å and 2.52 Å, with the largest Te-Fe-Te angle being 96.1°. The orbital-projected occupation numbers of the Fe-3$d$ orbitals for the NM ground state are calculated and shown in Fig. 3(b). Due to the breaking of $O_h$ symmetry, the degeneracy of the CF levels is removed. The occupation numbers of the five 3$d$ orbitals suggest an energy order from the lowest to the highest to be: $d_{x^2-y^2}$, $d_{yz}$, $d_{xz}$, $d_{z^2}$ and $d_{xy}$. The expected $e_g^0 t_{2g}^{\uparrow\downarrow\uparrow\downarrow\uparrow\downarrow}$ spin configuration, which is in the limits of both localized electrons and strictly octahedral CF, is not observed in marcasite $FeTe_2$. These two conditions are not met, due to the hybridization between Fe-$d$ and Te-$p$ orbitals, and the distortions of the $FeTe_6$ octahedral structure. Nonetheless, a preference for $t_{2g}$ orbital occupation can still be identified according to the total occupation number (4.4 for $t_{2g}$ and 1.66 for $e_g$).

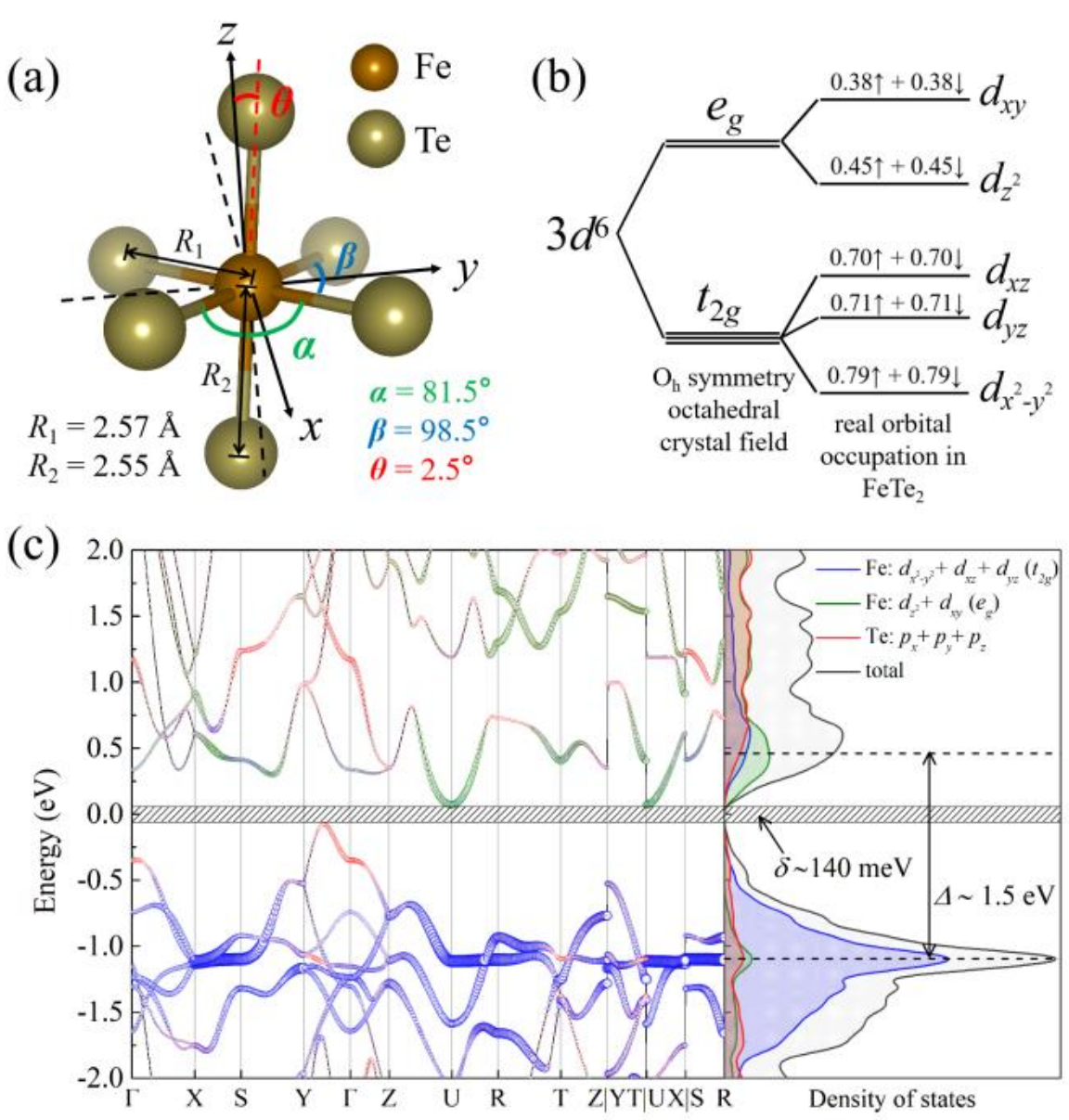

FIG. 3. (a) The distorted $FeTe_6$ octahedral structure in the marcasite $FeTe_2$ crystal. (b) The splitting of the 3$d$ orbitals in real $FeTe_2$ crystal environment and the spin occupation numbers of each 3$d$ orbital obtained by DFT calculations. (c) The electronic structure of the NM ground state of $FeTe_2$ with $U_{eff}$ = 1.2 eV. The contributions to the bands from Fe-$d$: $t_{2g}$, Fe-$d$: $e_g$, and Te-$p$ orbitals are represented by blue, green and red circles, respectively.

The corresponding electronic structure of the NM ground state is shown in Fig. 3(c). The orbital-resolved DOS shows slight hybridization between $t_{2g}$ and $e_g$ CF orbitals as expected in the distorted $FeTe_6$ structure. The CF splitting $\Delta$ is estimated to be 1.5 eV, which is well consistent with the simulated value of $U_{eff}$ at the NM-AFM phase transition point. With $U_{eff}$ being 1.2 eV, the indirect gap $\delta$ of $FeTe_2$ bulk is predicted to be 140 meV. Any magnetic ground states for $FeTe_2$ bulk have been excluded due to the non-gapped feature. Spin-orbit coupling (SOC) in $FeTe_2$ does not provide qualitative changes to the ground-state bands except for some reductions of the electronic degeneracy in the Brillouin Zone. The SOC bands of the NM ground state are shown in Section H of the SM [28]. In summary, it is demonstrated by the DFT calculations that the CF splitting in $FeTe_2$ bulk prevails over on-site Coulomb repulsion to stabilize a NM ground state.

### C. The magnetic feature on $FeTe_2$ surfaces

The CF splitting in $FeTe_2$ bulk is large enough to form the NM ground state as we have just clarified. However, our DFT calculations show spontaneous spin polarization from the outermost Fe atoms on the surfaces of the $FeTe_2$ bulk. As aperiodic boundaries, the type of the terminated atoms can be Fe, Te, or both of them (like on the 100 surface [18,19]). In this work, we simulate the 010 surfaces and 001 surfaces of $FeTe_2$. These crystal surfaces only include the Te-terminated (Te-T) surface or the Fe-terminated (Fe-T) surface. The slab models of the atomic structures are applied to simulate the surface properties as introduced in Section I of the SM [28]. In the following part of this paper, we only discuss the formation of localized moments on the 010 surfaces, since the physical mechanism for the spin-polarized 001 surfaces is the same as that for the 010 surfaces.

The spin-polarized Fe atoms on the surfaces of $FeTe_2$ tend to form 2D magnetic orders due to the superexchange interaction mediated by Te atoms. By computing the total energies of the magnetic structures, it is confirmed that the 2D AFM structures have the lowest energies for both the Te-T surfaces and the Fe-T surfaces. The magnetic anisotropy of the magnetic Fe atoms on the surfaces is also simulated. Notable magnetic anisotropy energies (MAEs) are recommended by our DFT calculations. The detailed data of the magnitude of localized magnetic moments, relative energies of the 2D magnetic structures and the MAEs are provided in Section J of the SM [28].

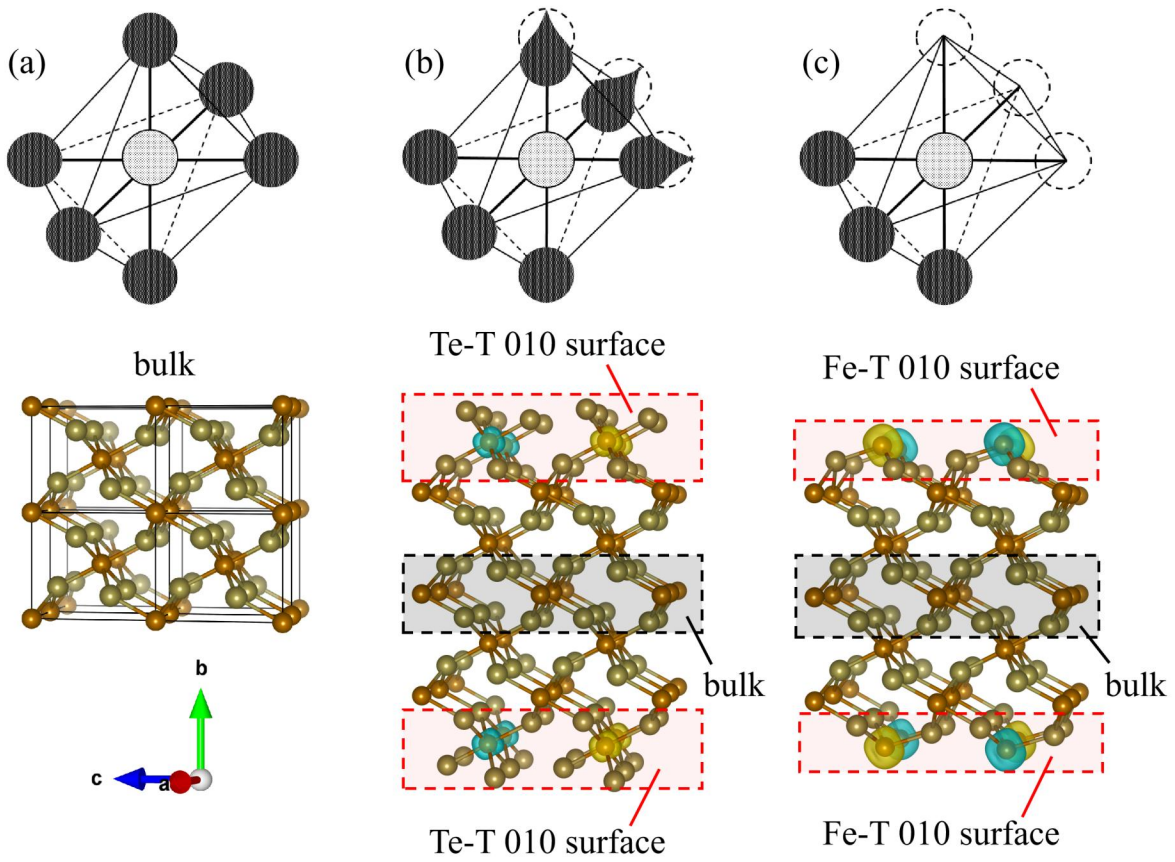

FIG. 4. The breaking patterns of the octahedral CF and the corresponding ground-state spin densities for (a): bulk, (b): Te-T 010 surface, and (c): Fe-T 010 surface of $FeTe_2$. The dark spheres and dark distorted spheres in the upper three sketches represent the charge densities of the bulk Te atoms and the outermost Te atoms, respectively. The dashed circles mark the missing charge densities from the Te atoms. The yellow and blue isosurfaces (0.02 spin/$Å^3$) in the lower three sketches represent the up spin density and the down spin density of the outermost Fe atoms, respectively.

We point out that the magnetic feature of Fe on the surfaces primarily originates from the breaking of the nearly octahedral CF formed by the six nearest Te ligands. The breaking patterns of the nearly octahedral CF and the corresponding spin densities which imply local magnetic moments, are shown in Fig. 4(a), (b), and (c) for Fe in the bulk, Fe on the Te-T 010 surface, and Fe on the Fe-T 010 surface, respectively. The corresponding localized spin magnetic moments are simulated to be 0 $\mu_B$, 1.28 $\mu_B$, and 2.80 $\mu_B$, respectively.

For Fe in the bulk, the six Te ligands have nearly equal charge densities. The octahedral CF thus possesses both the approximate symmetry of the $O_h$ point group and a large splitting $\Delta$, which stabilize the NM state. For Fe on the Te-T 010 surface, the three Te ligands on the surface experience a lack of attractive potential field from the vacuum side of the crystal. This different environment alters the charge densities of the three terminated Te atoms, thereby lowering the symmetry of the original octahedral CF for the central Fe. The CF splitting is thus reduced to $\Delta_1$ from $\Delta$, which also reduces the energy cost for higher-energy CF orbital occupation. As a result, a net spin moment on Fe can occur due to a relatively larger reduction of on-site Coulomb repulsive energy. For Fe on the Fe-T 010 surface, three of its Te ligands from the octahedral structure are removed. In this case, not only is the symmetry of the CF badly broken, but also the strength of the CF potential is weakened. The CF splitting reduces to $\Delta_2$ from $\Delta$, and $\Delta_2$ should be much smaller than $\Delta_1$ as well. Consequently, the net spin of Fe on the Fe-T surface is larger than that of Fe on the Te-T surface.

Given that the maximum net spin magnetic moment for Fe atoms can be 4 $\mu_B$, the unsaturated spin moments recommended by our DFT calculations imply residual CF strength for the surface Fe atoms. We exclude the contribution to the net spin from the variation in the valence of Fe atoms. A charge transfer analysis is given in Section K of the SM [28], and it is shown that the $3d^6$ electronic configuration of $Fe^{2+}$ is well maintained in bulk, Te-T surface and Fe-T surface of marcasite $FeTe_2$.

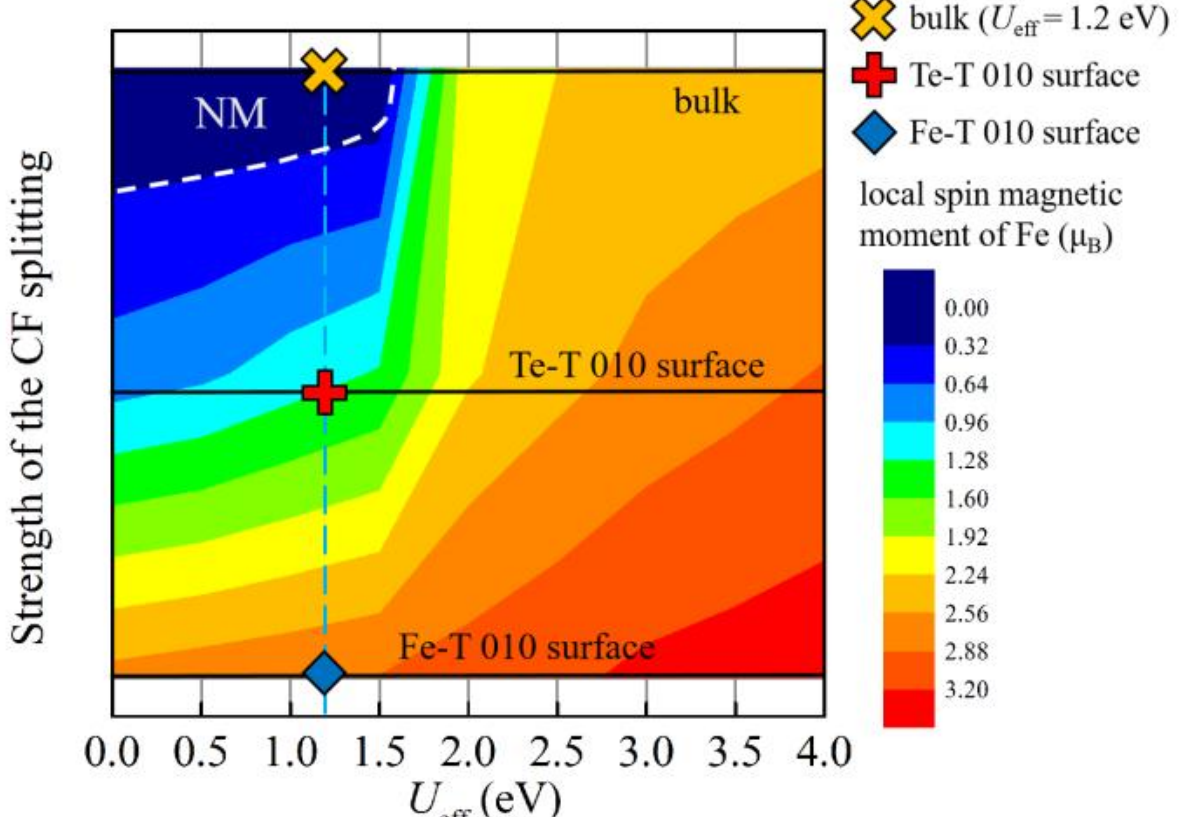

FIG. 5. The localized magnetic moment of Fe controlled by $U_{eff}$ and the strength of CF splitting. The NM phase of $FeTe_2$ is located at the top left corner of the phase diagram. The color distribution is achieved by interpolating the originally simulated points in the phase diagram. There are a total of three sampling points vertically (bulk, Te-T 010 surface, and Fe-T 010 surface) and nine sampling points horizontally (ranging from 0 to 4 eV with a step of 0.5 eV).

Fig. 5 depicts the simulated localized magnetic moment of Fe controlled by $U_{eff}$ and the strength of CF splitting. When $U_{eff}$ is small in bulk $FeTe_2$, and with the strong CF splitting, the localized moment of Fe is quenched. The NM-AFM phase transition occurs at $U_{eff}$ = 1.5 eV, beyond which the localized moment increases with increasing $U_{eff}$. When the CF splitting is weakened in $FeTe_2$, as on the Te-T 010 surface or Fe-T 010 surface, the localized moment can always be present and increase with the increasing $U_{eff}$. It should be noted that the applied PBE functional inherently incorporates a portion of the electronic correlation effect. This can provide enough Coulomb repulsive energy to facilitate the formation of localized moments on the surfaces, even in the absence of an explicit $U_{eff}$. Now it can be concluded that the presence of localized magnetic moments on the surfaces of $FeTe_2$ is due to the victory of on-site Coulomb repulsion in the competition with CF splitting.

## III. DISCUSSION

The ground state of marcasite $FeTe_2$ is confirmed to be NM in the bulk and AFM on the surfaces. This picture of marcasite $FeTe_2$ can explain the abnormal experimental observations, such as the unsaturated magnetic moment of Fe. In other words, the measurements of the total magnetization are not suitable for determining the localized magnetic moment of Fe without characterizing the shape and the size of the samples. Since the localized moments of most Fe atoms are quenched inside the bulk, this dilutes the estimated magnetization averaged to each magnetic Fe atom on the surfaces. Moreover, given that the breaking CF helps to form localized moments, we expect that grain boundaries, impurities [20] and defects [36,37] may bring various magnetic responses to $FeTe_2$ and experimentally cause deviations from typical magnetic systems. This highlights the significance of high-quality single-crystal samples in identifying the ground state of $FeTe_2$.

A recent experiment on quasi-2D marcasite $FeTe_2$ nanocrystal provides the thickness (about 10 nm) of their sample [25], which can be used to estimate the surface magnetic moment. Their measured magnetic moment is about 0.043 $\mu_B$ to 0.064 $\mu_B$ per Fe, showing a severely unsaturated feature. Adopting the surface-magnetism scheme, the estimated magnetic moment of the surface Fe atoms becomes about 0.74 $\mu_B$ to 1.60 $\mu_B$, which are in the same order of magnitude as Fe on Te-T surfaces predicted by our DFT calculations. As a small-gap semiconductor, the potential magnetic orders on its surfaces or one-dimensional edges may also bring about abundant in-gap states. The variable widths of the gap at the surfaces and edges of $FeTe_2$ have been detected by another STS study [18], as well as a DFT study [19]. The correlation between surface magnetism and in-gap states makes $FeTe_2$ a potential platform to explore exotic quantum states.

## IV. CONCLUSIONS

In this work, DFT+U calculations combined with scanning tunneling microscopy/spectroscopy are applied to study the magnetism of ground-state marcasite $FeTe_2$ single crystal. The competition between crystal field splitting and on-site Coulomb repulsion is proposed to judge the formation of localized magnetic moments in marcasite $FeTe_2$. It is recommended by the DFT calculations that the $FeTe_2$ bulk is NM while the $FeTe_2$ surfaces are magnetic.

This understanding of ground-state $FeTe_2$ also offers insights into explaining abnormal phenomena observed in magnetic measurements for other CF-Coulomb-competing systems.

## ACKNOWLEDGMENTS

This work is supported by the National Key R&D Program of China (2022YFA1403100, 2022YFA1403101 and 2022YFA1403103), National Natural Science Foundation of China (No. 12175023, 12275031, and 22372096), Shenzhen Polytechnic Research Fund (No. 6023310019K and 6022312037K), and Post-doctoral Later-stage Foundation Project of Shenzhen Polytechnic University (No. 6023271020K).